\documentclass[a4paper]{article} 

\usepackage{textcomp}
\usepackage{url}
\usepackage[latin1]{inputenc} 
\usepackage{multicol}
\usepackage{subfigure}
\usepackage{graphicx}
\usepackage{amsmath}
\usepackage{amssymb}
\usepackage{color}
\usepackage[thinqspace,squaren] {SIunits} 
\usepackage{txfonts}
\usepackage{ulem}
\usepackage{fancyhdr}

\newcommand{\Gstr}{G^\mathrm{\scriptscriptstyle{{ss}}}_\mathrm{\scriptscriptstyle{str}}(n,\alpha,f)}
\newcommand{\Gform}{G_\mathrm{form}(n,\alpha)}
\newcommand{\dxss}{\Delta x^\mathrm{\scriptscriptstyle{ss}}_l}

\newcommand{\smni}{^{\scriptscriptstyle{0}}_{\scriptscriptstyle{i}}}

\begin{document}

\title{Folding and unfolding of a triple-branch DNA molecule with four conformational states}
 
\author{\parbox{10cm}{\begin{center}{Sandra Engel$^{\rm a}$\thanks{ sandra.engel@uos.de}, 
Anna Alemany$^{\rm{b,c}}$, Nuria Forns$^{\rm{b,c}}$,\\ Philipp Maass$^{\rm a}$\thanks{philipp.maass@uos.de, phone: 0049-541-969-3460, fax: 0049-541-969-2351, {\tt{ http://www.statphys.uni-osnabrueck.de}}}, and Felix Ritort$^{\rm {b,c}}$ \thanks{fritort@gmail.com, phone: 0034-934035869, fax: 0034-934021149, \tt http://www.ffn.ub.es/ritort/index.html}\\
\vspace{6pt}     
\small{
    $^{\rm a}${\it{Fachbereich Physik, Universit\"at Osnabr\"uck,\\ Barbarastr. 7, 49076 Osnabr\"uck, Germany}}}\\
\small{
    $^{\rm b}${\it{Departament de F\'{i}sica Fonamental, Facultat de F\'{i}sica,\\ Universitat de Barcelona, Diagonal 647, 08028 Barcelona, Spain}}}\\
\small{
    $^{\rm c}${\it{CIBER-BBN Networking center on Bioengineering,\\ Biomaterials and Nanomedicine, Spain}}} }\end{center}   
}}

\date{}

\maketitle

 \vspace{-0.7cm}
\begin{abstract}
Single-molecule experiments provide new insights into biological
processes hitherto not accessible by measurements performed on bulk
systems. We report on a study of the kinetics of a triple-branch
DNA molecule with four conformational states by pulling
experiments with optical tweezers and theoretical modelling. Three
distinct force rips associated with different transitions between the
conformational states are observed in the folding and unfolding
trajectories.
By applying transition rate theory to a free energy
model of the molecule, probability distributions for the first rupture forces
of the different transitions are calculated. Good agreement of the
theoretical predictions with the experimental findings is achieved. Furthermore, due to our specific design of the molecule, we found a useful method to identify permanently frayed molecules by estimating the number of opened basepairs from the measured force jump values.\\

\begin{center}
optical tweezers, DNA\\ \end{center}

\begin{center}
\end{center}

\end{abstract}

\section{Introduction}

In recent years, single-molecule experiments became of great importance in biophysical research since progress in nano- and microscale manufacturing technologies facilitated the design of scientific instruments with sufficient sensitivity and precision to enable the controlled manipulation of individual molecules (for reviews, see, for example, \cite{Ritort2006, Hormeno2006}). 
In contrast to the traditionally used bulk assays, where individual biomolecular dynamics can get masked, single-molecule experiments provide new insights into the thermodynamics and kinetics of biophysical and biochemical processes hitherto not accessible. 
They complement standard spectroscopy and microscopy methods used in molecular biology and biochemistry and hence have to be regarded as an important source of additional information helping in the interpretation of biomolecular processes.
Furthermore, single-molecule experiments permit the measurement of small energies and the detection of large fluctuations. 

The manipulation of single molecules offers a powerful new tool in molecular and cellular biophysics allowing for the exploration of processes occuring inside the cell at an unprecedented level.
To instance just a few of the recently investigated biochemical processes: the transport of matter through pores or channels \cite{Melchionna2007,Meller2000,Ammenti2009}, 
interactions between DNA and proteins \cite{Leger1998}
or DNA and RNA \cite{Yin1995}, 
the motion of single-molecular motors \cite{Block2003,Carter2006,Wen2008}, 
DNA transcription and replication \cite{Wuite2000,Maier2000},
virus infection \cite{Smith2001,Dumont2006}, 
DNA condensation \cite{Ritort2006a}
and ATP generation \cite{Yasuda2001}.
In addition, the structure of biological networks \cite{Wagner2006}
and the viscoelastic and rheological properties of the DNA 
\cite{Smith1996,Wang1997,Dessinges2002,Mameren2009}
have been studied.

An important class of single-molecule experiments are performed with optical tweezers. By means of an optical trap generated by a focused laser beam, this useful technique renders it possible to exert forces on micron sized objects, achieving sub-piconewton and sub-nanometer resolution in force and extension, respectively.
Accordingly, one can study force-induced folding-unfolding dynamics and in this way get insight into corresponding processes in the cell and typical bond forces. Of particular interest is the unfolding of DNA molecules, where the hydrogen bonds between the complementary base pairs (bps) are disrupted. This so-called unzipping is connected to the DNA replication mechanism.

An interesting field of biophysical studies is the investigation of junctions in molecules since they present manifold ways to interact with other substances, for instance cations. Three-way junctions are especially interesting because metal ions such as magnesium can bind to them and alter the tertiary structure. Here a first step of such a study is presented where we investigate a molecule with a three-way junction alone, without cation binding.

Many of the single-molecule experiments so far focused on molecules with a relatively simple free energy lanscape (FEL) exhibiting just two states, a folded and an unfolded one, or including an additional misfolding state, leading to different kinetic pathways.
In this work we will consider a richer situation, where metastable states as intermediates occur during the folding-unfolding route. In this context, we will address the following key questions:
\begin{enumerate}
\item[(i)] Can a corresponding molecule with such intermediate states be designed on the basis of a suitable model for a FEL?
\item[(ii)] Is it possible to observe the intermediate states by perfoming pulling experiments with optical tweezers?
\item[(iii)] Can phenomenological Bell-Evans kinetic models be applied to describe the folding-unfolding processes including intermediate states? In particular, when validating the kinetic theory against the experimental results, how do the first rupture force distributions compare with the ones predicted by the theory? 
\end{enumerate}

A further important aspect that we looked at in some detail is the heterogenity of molecular folding-unfolding behaviour that we observed in the experiments. By measuring force-distance curves (FDCs) of several molecules we classified them into different reproducible patterns. This leads
to a useful method to identify irreversible molecular fraying, a phenomenon which is often observed in single-molecule studies.

\section{Description of the experiments}\label{sec: exp}

\begin{figure}[htbp]
\begin{center}
\includegraphics[scale=0.4]{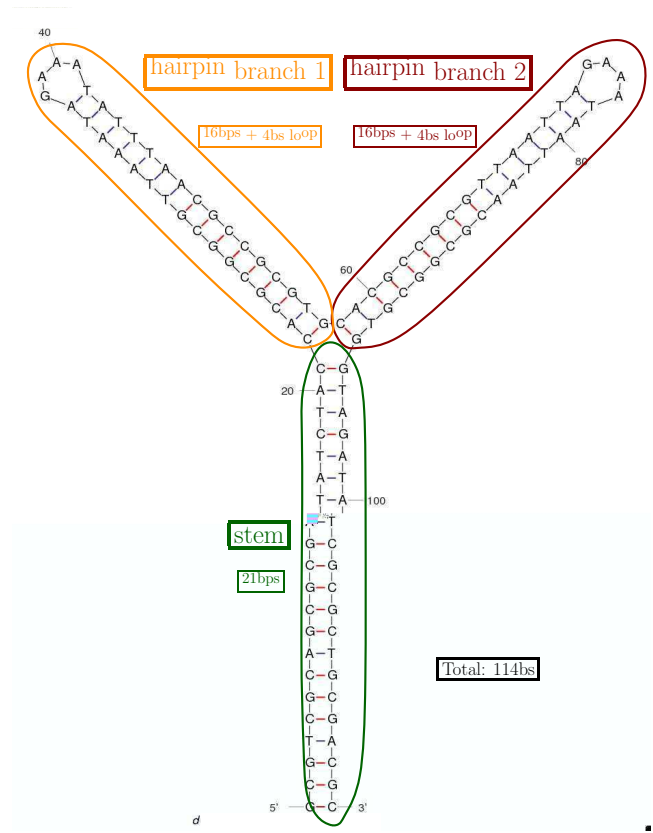}
\end{center}
\caption{Structure of the triple-branch DNA molecule.}
\label{fig: 3helixmol}
\end{figure}

Based on Mfold folding predictions \cite{SantaLucia1998,Zuker2003} and taking FEL considerations into account (see sec.~\ref{sec: theory}), we designed and synthesised a DNA molecule which is composed of three parts and hence referred to as triple-branch molecule. It consists of a stem as introduced in \cite{Mossa2009a} with $\unit{21}{bps}$ and two nearly identical hairpin branches which are formed of 16 bps and a loop with four bases, thus comprising a total number of \unit{114}{bases}, see fig.~\ref{fig: 3helixmol}. To avoid misfolding, the second hairpin branch differs at two positions from the first one. 

\begin{figure}[htb]
\begin{center}
\includegraphics[scale=0.91]{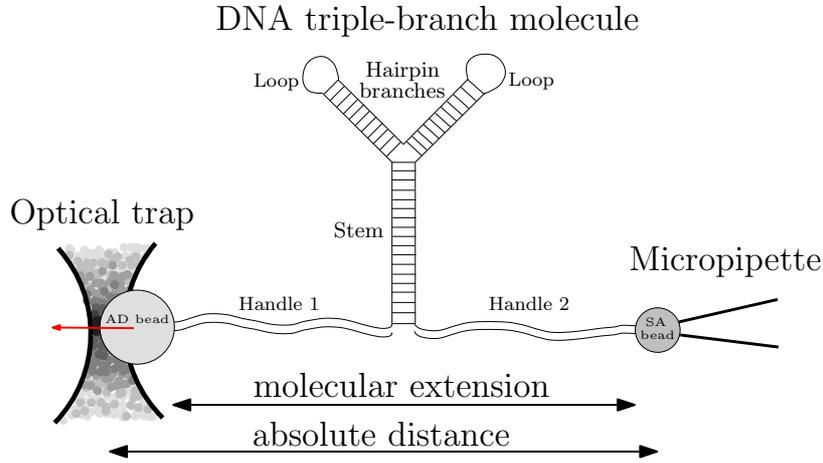}
\end{center}
\caption{Sketch of the experimental setup. In our experiments we measure the relative distance $X$ rather than its absolute value.}
\label{fig: model_molecular_setup}
\end{figure}

The triple-branch molecule is inserted between two identical short double-stranded DNA (dsDNA) handles of $\unit{29}{bps}$ each \cite{Forns2011}, leading to a total number of $\unit{172}{bases}$, corresponding to a total contour length in the unfolded state of about $\unit{100}{\nano\meter}$.
Each of these polymer spacers is chemically linked to a bead. One of the beads is retained with the help of a pipette via air suction, the other one is optically trapped in a laser focus \cite{Smith2003}.
On the $5'$ end of the DNA molecule, biotin is attached to enable a connection with a streptavidin-coated bead (SA bead) whose diameter is \unit{1.8}{\micro\metre}. Biotin is a vitamin which establishes a strong linkage to the proteins avidin and streptavidin. The $3'$ end is modified with the antigen digoxigenin, able to interact with an antidigoxigenin-coated bead (AD bead). The latter has a diameter of \unit{3.0}{\micro\metre}.
Figure \ref{fig: model_molecular_setup} shows the different components of the molecular construct, that is to say the triple-branch DNA molecule, the handles and the beads, captured in optical trap and micropipette, respectively. Note that it is not a true-to-scale representation.

The pulling experiments are carried out with a miniaturised dual-beam laser optical tweezers apparatus \cite{Huguet2010} at room temperature ($\simeq$ 25 \textcelsius) and at salt concentration of 1 M NaCl aqueous buffer with neutral pH (7.5) stabilised by Tris HCl and 1 M EDTA.
The dual-beam optical tweezers collect data at \unit{4}{\kilo\hertz} and can operate with a feedback rate of \unit{1}{\kilo\hertz}. Spatial resolution constitutes \unit{0.5}{\nano\metre} with a maximal distance range of $\simeq \unit{10}{\micro\metre}$. Forces up to \unit{100}{\pico\newton} can be achieved, whereas the force resolution is \unit{0.05}{\pico\newton}.

Optical tweezer pulling experiments permit the measurement of the force $f$ as well as the total distance $X$ between the centre of the optical trap and the tip of the micropipette, see fig.~\ref{fig: model_molecular_setup}.
In the experiment we vary the trap-pipette distance $X(t)$ with a constant speed $v=dX/dt$ in the range of $\unit{45}{\nano\metre\per\second}$ to $\unit{200}{\nano\metre\per\second}$, which corresponds to a constant average loading rate $r$ of $\unit{3.0}{\pico\newton\per\second}$ and $\unit{13.4}{\pico\newton\per\second}$ in between rip events, respectively. 
The experiment consists of loading cycles which in turn are divided into an unfolding part (during "pulling") and a folding part (during "pushing").
The loading cycles are repeated as long as the tether connection is unbroken. Otherwise a new connection has to be established, possibly a new molecule must be searched and linked to a new bead. 
In sec.~\ref{sec: patterns} we will discuss different patterns found in the measured curves and present a detailed analysis of two representative molecules. We chose them among seven molecules exhibiting the first and among five molecules featuring the second pattern. For each molecule we recorded, on average, approximately 50 cycles. The pulling speeds, ranging from 45 to $\unit{200}{\nano\metre\per\second}$, influence the experimental results only weekly due to a logarithmic dependence of the first rupture force with the speed.
We found compatible data for sets of similar molecules. 
In the theoretical analysis of the data in sec.~\ref{sec: theory}, we concentrate on the largest set of 82 loading cycles for the molecule at a speed of  $\unit{200}{\nano\metre\per\second}$. 
The study of the above mentioned second molecule comprises 55 cycles.

\section{Analysis of unfolding and folding trajectories}

\begin{figure}[htb]
\begin{center}
\includegraphics[scale=0.95]{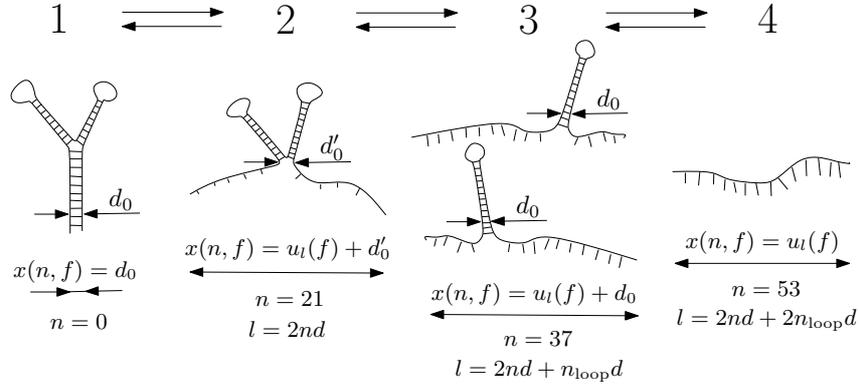} 
\end{center}
\caption{The four stable or metastable states of the triple-branch molecule: 1 - folded molecule, 2 - unfolded stem, 3 - stem and one hairpin branch unfolded, 4 - unfolded molecule. The values of $x(n,f)$ refer to the end-to-end distance given in eq.~(\ref{eq: eq_dist}) with the number $n$ of opened bps corresponding to the molecular construct shown in fig.~\ref{fig: 3helixmol} and the contour length $l$ according to eq.~(\ref{eq: contour}).}
\label{fig: states 3bmol}
\end{figure}

Based on the design of the triple-branch molecule, we have to distinguish between four conformational states (see fig.~\ref{fig: states 3bmol}):
\begin{enumerate}
\item a completely folded molecule.
\item a completely unfolded stem with the hairpin branches still folded.
\item stem and either hairpin branch 1 or 2 are completely unfolded.
\item a completely unfolded molecule. 
\end{enumerate}  

\begin{figure}[htb]
\begin{center}
\subfigure[]
{\label{fig: u_molA}
\includegraphics[scale=0.3,angle=-90]{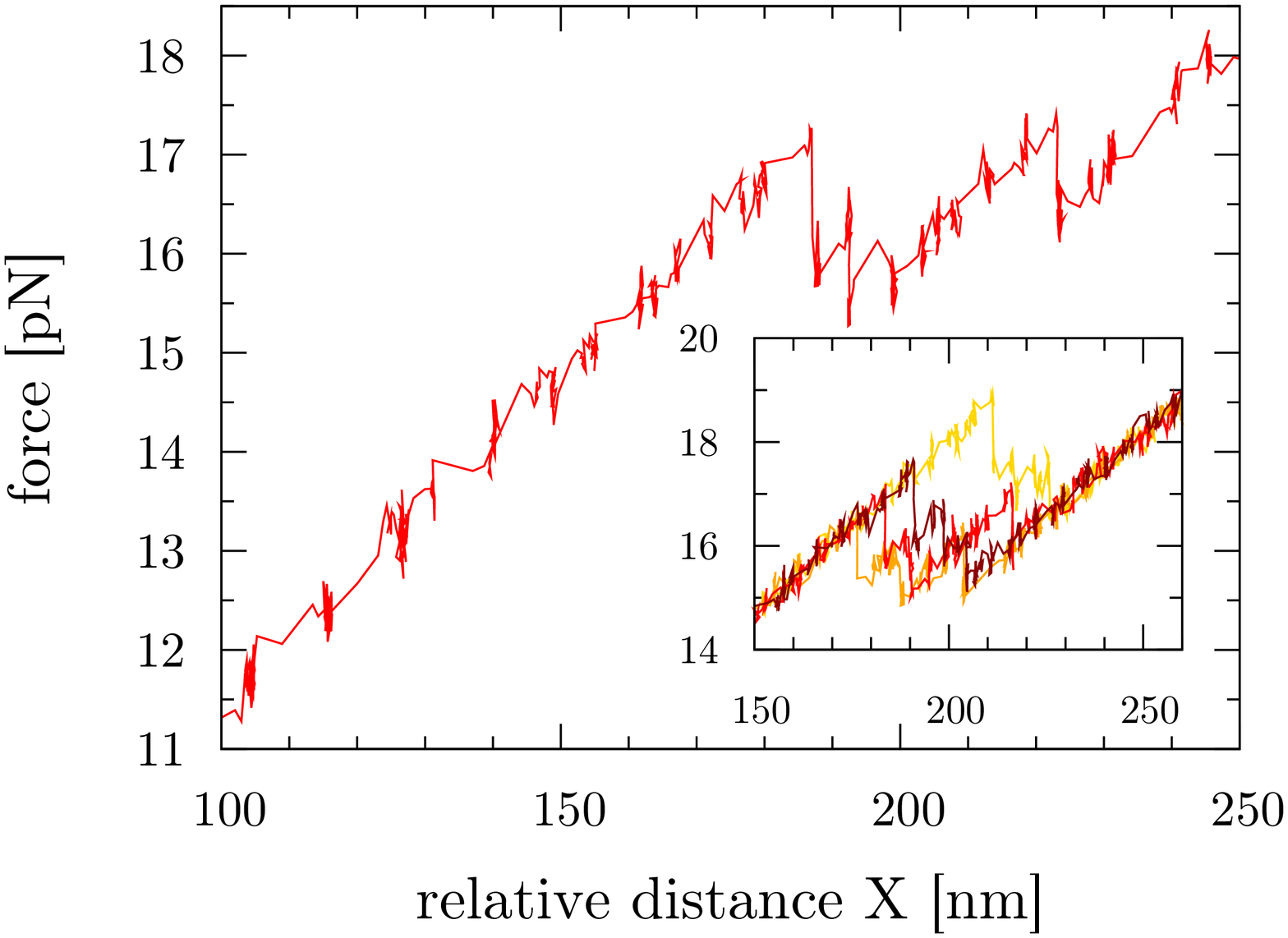}}   
\subfigure[]
{\label{fig: uf_molB}
\includegraphics[scale=0.3]{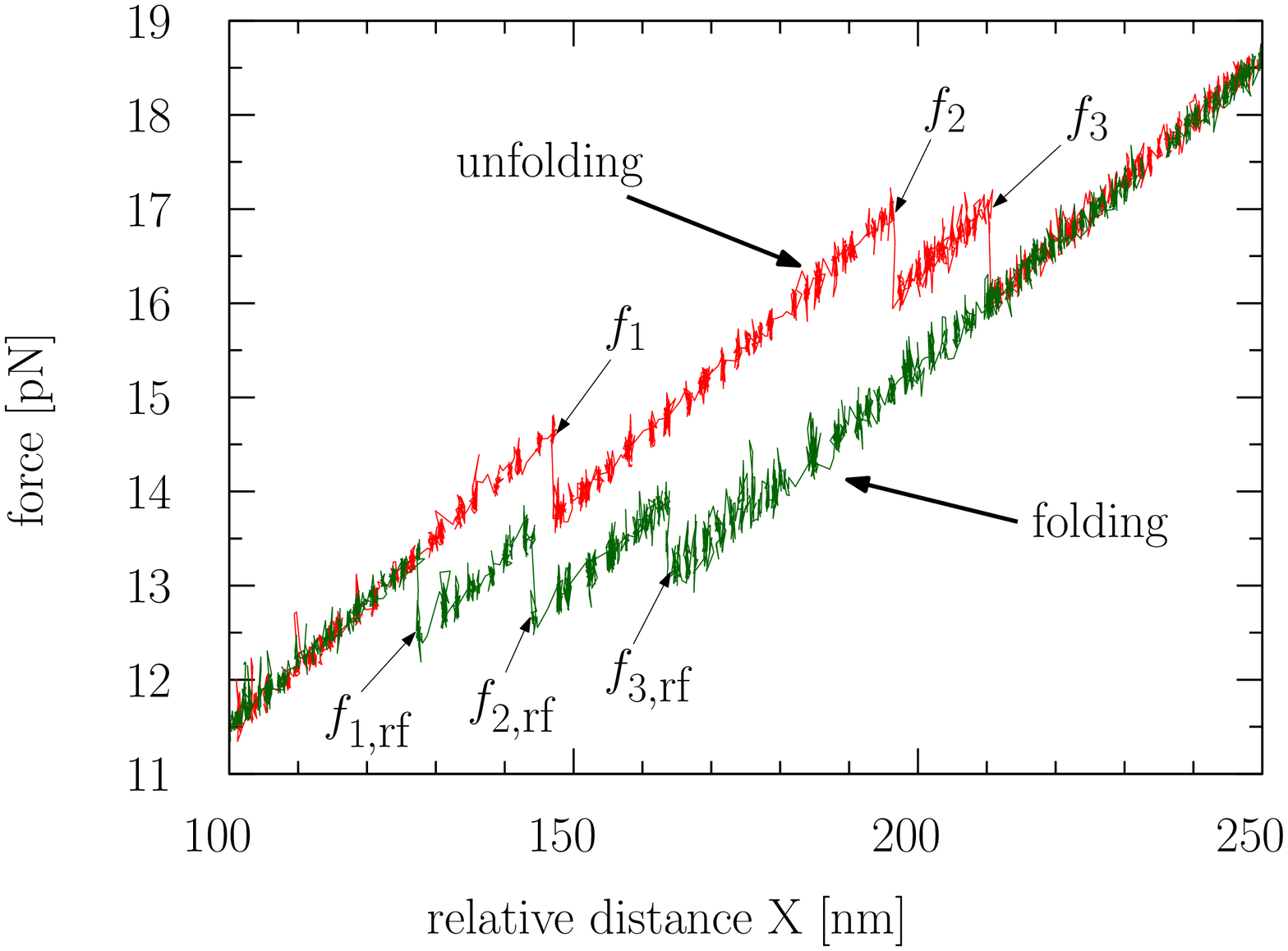}}
\end{center}
\caption{Force as a function of change in the trap-pipette distance (a) for one typical unfolding trajectory for the first investigated molecule and (b) for one typical unfolding and folding trajectory for the second investigated triple-branch molecule. Indicated are the three first rupture forces belonging to the transitions 1 ($f_1$), 2 ($f_2$) and 3 ($f_3$) and the respective refolding forces, labelled $f_{i,\textrm{rf}}$. The inset in (a) shows four further unfolding curves for the first investigated molecule.}
\label{fig: exp_results}
\end{figure}

Figure \ref{fig: u_molA} displays a typical unfolding route and fig.~\ref{fig: uf_molB} a typical unfolding and folding trajectory in form of a FDC, which in fact records the evolution of the force as a function of time and relative trap position $X$\footnote{ 
In our experiments we measure the relative distance between trap and pipette, $X$, rather than the absolute value.
The force $f$ exerted on the molecular construct leads to a displacement $f / k_{\rm{b}}$ of the bead in the optical trap, where $k_{\rm{b}} = \unit{0.08}{\pico\newton\per\nano\metre}$ is the rigidity of the trap. Hence, the relative distance $X$ is related to the relative molecular extension $x_{\rm{m}}$ (see fig.\ \ref{fig: model_molecular_setup}) by $x_{\rm{m}} = X - f / k_{\rm{b}}$.}.
With rising $X$, the force $f$ first increases almost linearly according to an elastic response of the DNA handles, which consist of dsDNA and are stable over the whole range of forces where the unfolding / folding of the triple-branch molecule takes place. The overstretching transition of the linkers, typically at $\unit{65}{\pico\newton}$, lies much above the forces we explore.
At a first rupture force $f_1$ a sudden decrease ("jump") $\Delta f_1$ occurs, which is caused by the unfolding of the stem.  
This unfolding goes along with an abrupt change in the length of the molecule when single-stranded DNA (ssDNA) is released. As a consequence, the bead in the optical trap moves towards the centre of the trap, visible as the force drops. 
Following the jump $\Delta f_1$, there is again a linear increase up to the next force rip at a first rupture force $f_2$, where one of the hairpin branches unfolds, which in turn leads to the force jump $\Delta f_2$. Eventually, the second hairpin branch unzips at a first rupture force $f_3$ with a jump $\Delta f_3$. The linear regime following this last force rip corresponds to the stretching of the whole molecular construct including handles and the already unfolded triple-branch molecule.

Upon decreasing $X$ from the completely unfolded state 4, the force first follows closely the corresponding unfolding part of the trajectory. However, a backward transition does not occur at $(f_3-\Delta f_3) \simeq \unit{16.2}{\pico\newton}$, but at a considerably lower value of $\unit{(14.0 \pm 0.8)}{\pico\newton}$, cp.~fig.~\ref{fig: uf_molB}, hence manifesting a hysteresis effect. Moreover, an investigation of a larger number of folding trajectories reveals that the bases of the hairpin branches do not always pair conjointly in well-defined events during a short time interval. In contrast, the corresponding force rips during unfolding indicate a cooperative behaviour of the biomolecule, where the breakage of all hydrogen bonds stabilising the DNA structure happens almost simultaneously. The second transition takes place at $\unit{(13.3 \pm 0.8)}{\pico\newton}$ instead of $(f_2-\Delta f_2) \simeq \unit{15.5}{\pico\newton}$.
After both hairpin branches refolded, one can identify another sharp transition to the folded state 1 around $\unit{(11.2 \pm 0.8)}{\pico\newton}$, which is again considerably lower than $(f_1-\Delta f_1) \simeq \unit{13.7}{\pico\newton}$.

During unfolding, one can assume that the breakage of hydrogen bonds follows the sequence of the molecular construct. This makes it useful to introduce the number $n$ of broken bonds as state variable and to calculate a FEL as function of this variable (see sec.~\ref{sec: theory}).  
The refolding of the hairpin branches in the folding trajectories exhibit less sharp transitions, see fig.\ \ref{fig: uf_molB}. During folding, in particular at the beginning in the unfolded state, a huge number of secondary structures can be found which implies that the kinetic pathways are less predefined and accordingly, the transitions get smeared out. With respect to a theoretical treatment, moreover, a description in terms of the simple state variable $n$ becomes unlikely to be sufficient. In a refined analysis, many more configurations should have to be included as relevant states in a coarse-grained description \cite{Manosas2008}. Such refined analysis, however, goes beyond the scope of this work amd we therefore concentrate on the unfolding process in the following.

\begin{figure}[htb]
\begin{center}
\subfigure[]
{\label{fig: fr_molA}
\includegraphics[scale=0.23,angle=-90]{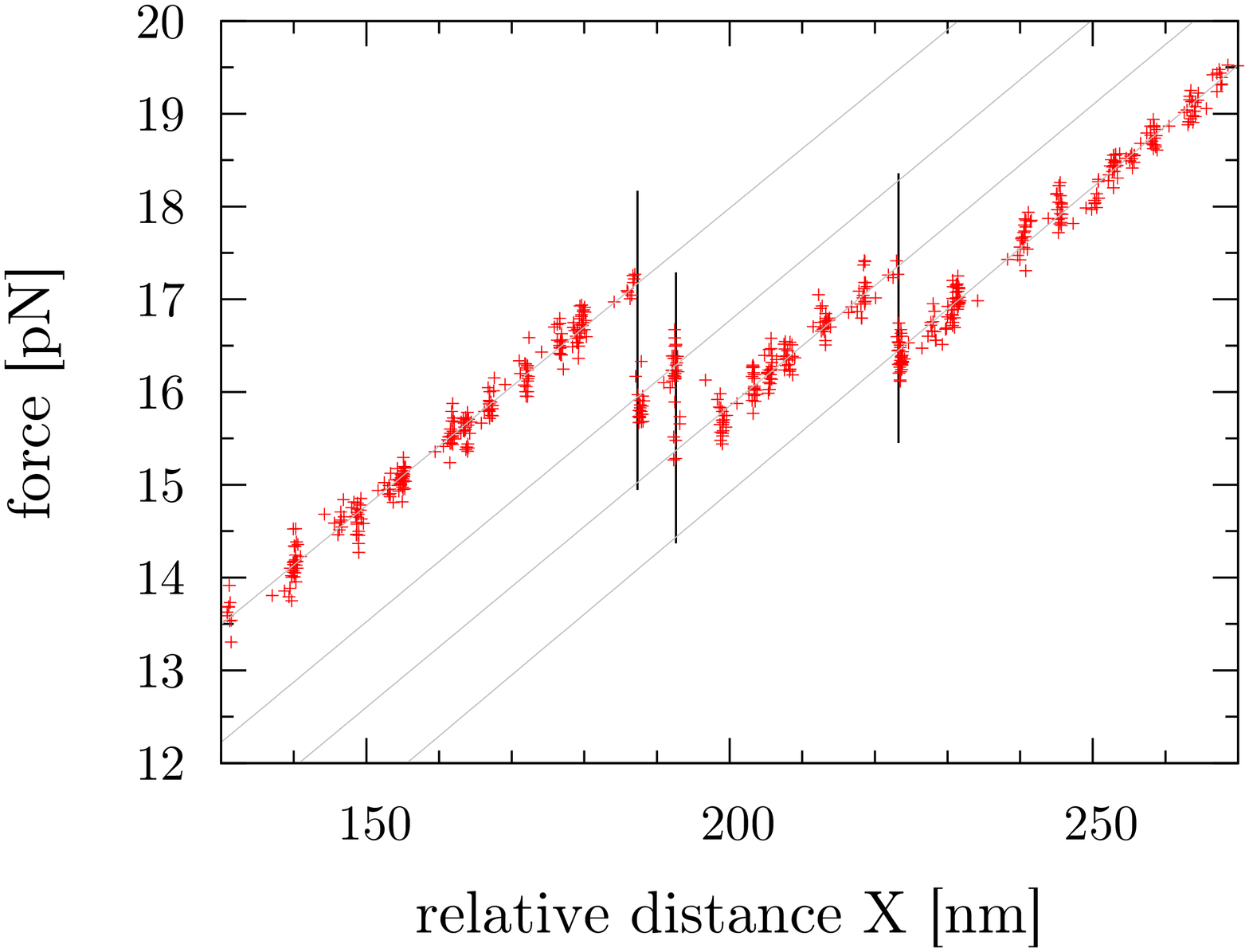}} 
\subfigure[]
{\label{fig: fr_molB}
\includegraphics[scale=0.23,angle=-90]{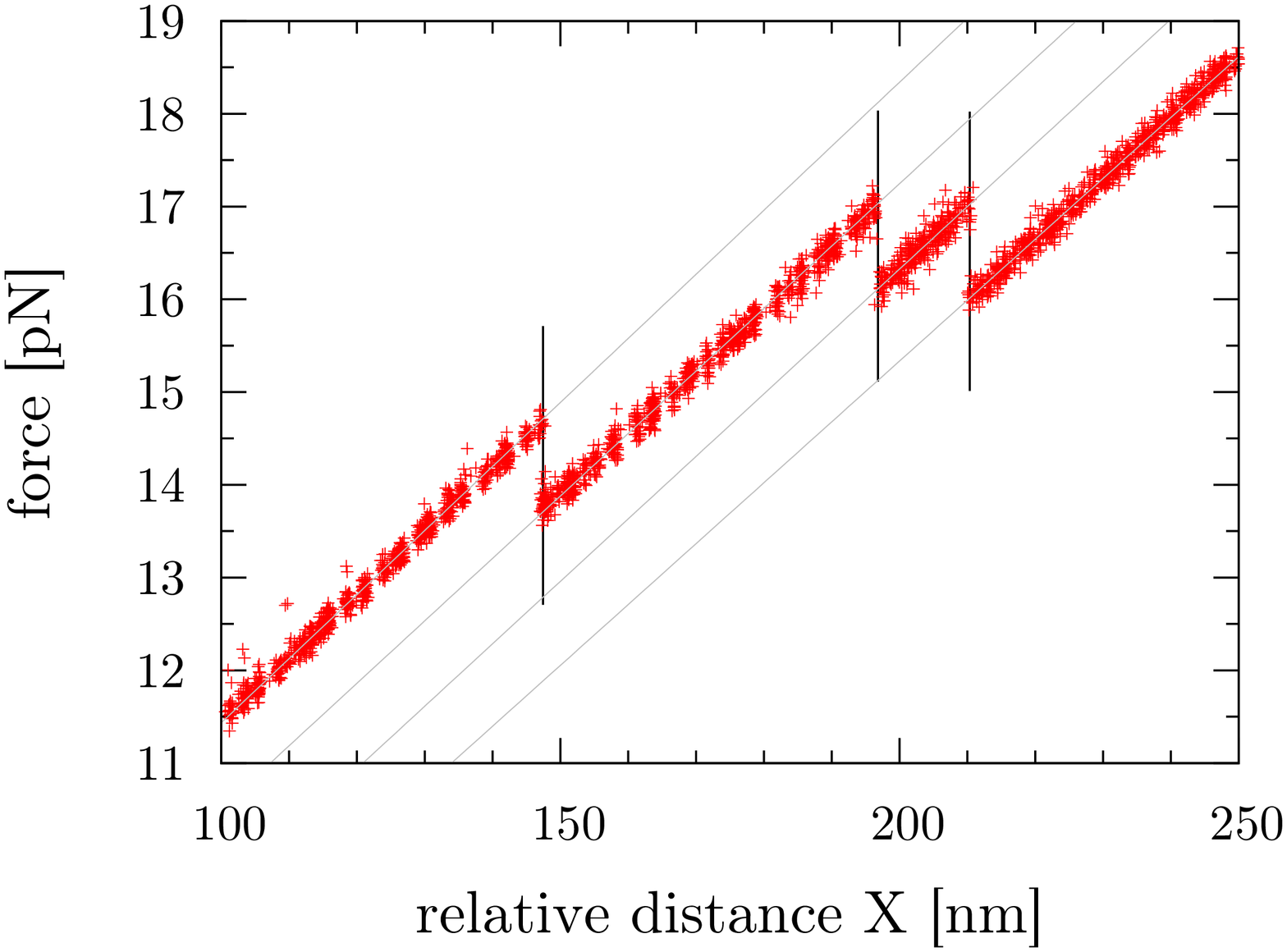}} 
\end{center}
\caption{The first rupture forces and force jump values of the unfolding trajectories shown in fig.~\ref{fig: exp_results} are extracted as indicated here from the intersection of the fitted grey lines and the transitions (vertical lines). In part (a) the procedure is shown for the molecule used in the theoretical analysis in sec.~\ref{sec: theory} and in (b) for the molecule with permanent fraying behaviour.}
\label{fig: find_rips}
\end{figure}

There are plenty of possible ways to analyse the FDCs in order to find out the first rupture forces and the force jumps.
In our procedure we arranged the normalised data, i.e. the relative distance $X$ and the force $f$, in windows of a certain size of data points. For all consecutive windows we then calculated the slope of the considered data points, the span, i.e. the maximum distance between the lowest and the highest force value, and the mean of the relative distance $\overline{X_j}$ as a moving average.
Transitions between the conformational states take place where the slope is minimal and the span maximal under the condition that an appropriate number of contiguous windows is connected. Having found the $\overline{X_j}$ of the three force rips, slope and axis intercept are calculated by linear regression for each conformational state. One can now easily calculate the first rupture forces as the intersection points with the four fitted lines and extract the force jump values, as exemplified in fig.~\ref{fig: find_rips} for both molecules whose unfolding trajectories were depicted in figs.~\ref{fig: u_molA} and \ref{fig: uf_molB}, respectively.
Since the data acquisition rate is constant, at higher pulling speeds less data points are collected and therefore the values of the analysis are broader distributed.
The three first rupture forces $f_i$ as well as the jumps $\Delta f_i$ are subject to stochastic fluctuations, as can be seen in fig.~\ref{fig: u_molA} where we show four unfolding trajectories belonging to different pulling cycles of the same molecule in the inset.
An analysis revealed that the fluctuations of the force jumps $\Delta f_i$ are about ten times smaller than the fluctuations of the $f_i$. Accordingly, in sec.~\ref{sec: theory}, we will disregard the fluctuations in the $\Delta f_i$ and use only their averages $\overline{\Delta f_i}$ that will be discussed in more detail in the following section.

\section{Different unfolding patterns related to the number of opened bps}\label{sec: patterns}

\begin{figure}[htb]
\begin{center}
\subfigure[]
{\label{fig: histo_frf_fj_molA}\includegraphics[scale=0.3,angle=-90] 
{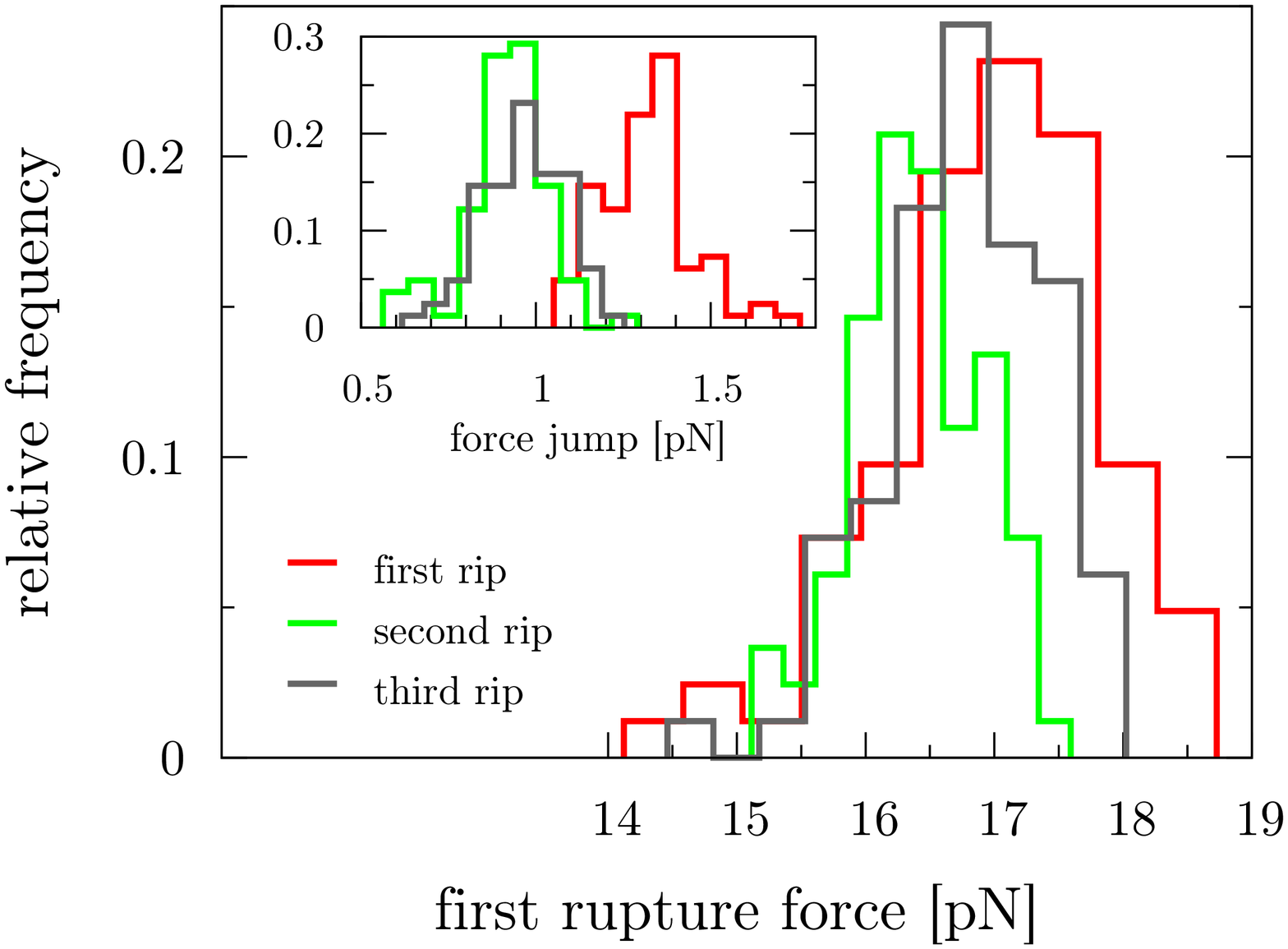}}
\subfigure[]
{\label{fig: histo_frf_fj_molB}\includegraphics[scale=0.3,angle=-90] 
{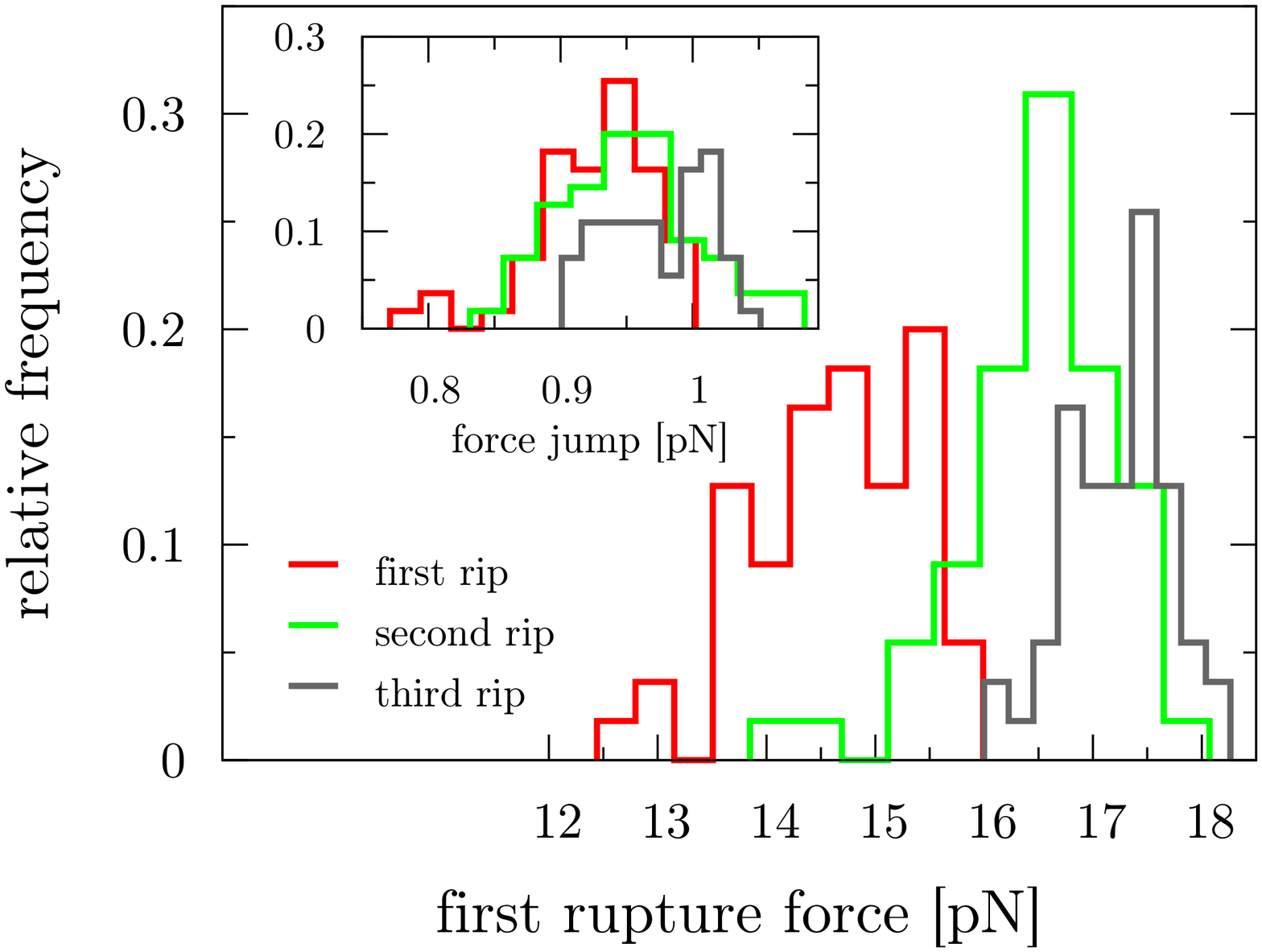}}
\end{center}
\caption{Histograms of the three first rupture forces during unfolding for two representative triple-branch molecules: (a) the one used in the theoretical analysis in sec.~\ref{sec: theory} and (b) the one exhibiting permanent fraying behaviour. Note that the average force value of the first rip has decreased in (b) as compared to (a). The insets show the corresponding histograms for the force jump values.}
\label{fig: frf_fj}
\end{figure}

Applying the above mentioned procedure to analyse the experimental data of several molecules, we found two predominating patterns in the unfolding trajectories which are reflected in the distributions of the first rupture forces $f_i$ as follows. In the first pattern, see fig.~\ref{fig: histo_frf_fj_molA}, these
distributions have a similar shape for all three force rips. The histograms indicate the existence of one maximum slightly below $\unit{17}{\pico\newton}$. 
The mean values for the three first rupture forces are $\overline{f_1} = \unit{(17.0 \pm 0.9)}{\pico\newton}$, $\overline{f_2} = \unit{(16.4 \pm 0.6)}{\pico\newton}$ and $\overline{f_3} = \unit{(16.8 \pm 0.7)}{\pico\newton}$.
In contrast, in a second pattern we detected a strikingly lower value for the first rupture force of the first rip  $\overline{f_1} = \unit{(14.6 \pm 0.8)}{\pico\newton}$, as depicted in  fig.~\ref{fig: histo_frf_fj_molB}, whereas the other two rip forces lie basically in the same range of 16 to $\unit{18}{\pico\newton}$. As in the former case, the second rip tends to have a slightly smaller first rupture force, $\overline{f_2} = \unit{(16.5 \pm 0.8)}{\pico\newton}$, than the third rip, $\overline{f_3} = \unit{(17.2 \pm 0.5)}{\pico\newton}$.

The small $\overline{f_1}$ observed in the second molecule suggests the occurrence of permanent molecular fraying. Obviously less force is needed to unfold the stem than typically, cp. fig.~\ref{fig: u_molA} with \ref{fig: uf_molB}, since some bps at its basis are partly or completely melted. In other words, during folding, this molecule does not reach an entirely folded state but some bps of the stem next to the handles remain irreversibly and permanently open.  
This phenomenon has been observed previously in several pulling experiments \cite{Huguet2010,Woodside2006a}.
A possible reason for irreversible fraying is the formation of reactive oxidative species due to the impact of the laser light of the optical trap leading to a degradation of the DNA bases \cite{Landry2009}.
The so generated singlet oxygens are known to oxidise certain nucleic acids, such as guanine and thymine, irreversibly. This could explain our observation that once a molecule shows fraying it does not change back again to normal behaviour.
Due to the fact that we work with polystyrene microspheres which are more prone to photodamage than the DNA bases themselves, their wide ranging interaction with the bases might be reduced replacing polystyrene by silica beads, which exhibit considerably minor irreversible oxidative damage. It would be very interesting to carry out such experiments.

The insets in figs.~\ref{fig: histo_frf_fj_molA} and \ref{fig: histo_frf_fj_molB} depict the corresponding force jump distributions of both molecules. While the first molecule possesses a large first force jump value of $\overline{{\Delta f_1}} = \unit{(1.3 \pm 0.2)}{\pico\newton}$ and two smaller force jumps at the second and third rip of $\overline{{\Delta f_2}} = \unit{(0.9 \pm 0.2)}{\pico\newton}$ and $\overline{{\Delta f_3}} = \unit{(1.0 \pm 0.2)}{\pico\newton}$, the frayed molecule features three force jumps of approximately the same value, i.e. $\overline{{\Delta f_1}} = \unit{(0.93 \pm 0.05)}{\pico\newton}$, $\overline{{\Delta f_2}} = \unit{(0.95 \pm 0.06)}{\pico\newton}$ and $\overline{{\Delta f_3}} = \unit{(0.97 \pm 0.04)}{\pico\newton}$, respectively.
This illustrates clearly the influence of irreversible fraying in the latter case since, roughly estimated, the same number of bps is expected to open in all three rips which should not be the case in an entirely folded molecule.

During a force rip, the relative distance $X$ is constant and thus $\Delta X = 0$. Therefore the change in the relative molecular extension,
\begin{equation} 
\Delta x_m = \Delta X - \Delta f / k_{\rm{eff}} = \Delta x(n,f),
\end{equation} 
is only related to the force jump $\Delta f$ and the  combined stiffness of bead and handles $k_{\rm{eff}}$, given by $1/k_{\rm{eff}} = 1 / k_{\rm{b}} + 1 / k_{\rm{h}}$, where $k_{\rm{b}}$ is the trap stiffness and $k_{\rm{h}}$ the rigidity of the handles, respectively.
With the help of a linear least squares fit, the effective stiffness  $k_{\rm{eff}}$ of the molecular construct is extracted
from the average slope of the FDCs and amounts to $\unit{(0.067 \pm 0.003)}{\pico\newton\per\nano\metre}$. Note that, as expected, $k_{\rm{eff}}$ is smaller than $k_{\rm{b}}$ ($\simeq \unit{0.08}{\pico\newton\per\nano\metre}$).

For a certain force value $f$ and assuming an elastic model for the released ssDNA, the number $n$ of opened bps is related univocally to the equilibrium end-to-end distance of the DNA molecule $x(n,f)$, whereas its change $\Delta x(n,f)$, in turn, equals $\Delta x_m$. Using this relation, it is now possible to estimate the change in the number $\Delta n_i$ of bps which are opened sequentially during each force rip. Considering only the configurations of the four conformational states, the equilibrium end-to-end distance $x(n,f)$ can be decomposed into two parts, cp.~fig.~\ref{fig: states 3bmol}. The first part, the elongation $u_l(f)$ of the mean end-to-end distance of the ssDNA  along the force direction, accounts for the ideal elastic response of the ssDNA, where $l$ is the contour length. The second part contains the contribution of the diameter of stem and hairpin branches, respectively. Acccordingly,
\begin{equation}
x(n,f) = u_l(f) + \left\{
\begin{array}{rl}
0,& n= $\unit{21+16+16}{bps}$\\
d_0,& n= 0 \text{ and } $\unit{21+16}{bps}$\\
d_0',&n=$\unit{21}{bps}$
\label{eq: eq_dist}
\end{array}, \right .
\end{equation}
where $d_0\simeq\unit{2}{\nano\metre}$, in accordance with the diameter of the B-DNA helix. The exact value of the diameter contribution of both hairpin branches $d_0'$, when the stem is unfolded, depends on the orientation of the branches.
We set $d_0' = 2d_0$ as a working value\footnote{Due to this simplification, the $\Delta n_1$ of the first rip is likely to be slightly underestimated and the second rip's $\Delta n_2$ overestimated. However, it will not affect the change in the total number $\Delta n_{\rm{tot}}$ of opened bps since we consider the change of $x(n,f)$, and the $d_0'$ contributions will cancel each other out.}.

Regarding the contour length $l$, which depends, amongst others (cp. sec.~\ref{sec: theory}), on the number $n$ of opened bps, and considering again solely the configurations of the four conformational states, one gets
\begin{equation}
l = 2nd + \left\{
\begin{array}{rl}
0,& n= 0 \text{ and } $\unit{21}{bps}$\\
n_{\rm{loop}}d,& n= $\unit{21+16}{bps}$\\
2n_{\rm{loop}}d,&n=$\unit{21 + 16 + 16}{bps}$
\label{eq: contour}
\end{array}, \right .
\end{equation}
where the interphosphate distance $d$ is taken to be $\unit{0.59}{nm/base}$ and $n_{\rm{loop}} = 4$ is the number of bases per loop. For every force rip $f_i$ the corresponding number of opened bps $n_i$ is calculated separately by considering the differences of $l$ between the states.

Different types of models can be used to calculate $u_l(f)$. Prominent examples borrowed from polymer physics are the freely jointed chain (FJC) and the worm-like chain (WLC) model. According to ref.~\cite{Smith1996}, the FJC model includes an extra term, leading to the expression
\begin{equation}
u_l(f)=l\left(1+\frac{f}{Y}\right)
\left[\coth\left(\frac{bf}{k_\mathrm{B}T}\right)-\frac{k_\mathrm{B}T}{bf}\right]\,. 
\label{eq: FJC}
\end{equation}
Here $Y$ denotes the Young modulus, $b$ is the Kuhn length, $k_\mathrm{B}$ the Boltzmann constant and $T$ the temperature. Typical values of the Kuhn length and the Young modulus under working conditions of $T\simeq 25$
\textcelsius $ $ and $\unit{1}{M}$ NaCl concentration are $b=\unit{1.42}{\nano\meter}$ 
and $Y=\unit{812}{\pico\newton}$ \cite{Smith1996} or, as published recently, $b=\unit{1.15}{\nano\meter}$ 
and $Y= \infty$ \cite{Huguet2010}, respectively.
In the WLC model \cite{Bustamante1994a}, the force $f(u_l)$, due to an elongation $u_l$, is given by
\begin{equation}
f (u_l)= \frac{k_{\rm{B}}T}{P}\left[\frac{1}{4\left(1-u_l/l \right)^2} -\frac{1}{4} + \frac{u_l}{l}\right],
\label{eq: WLC} 
\end{equation}
and to obtain $u_l(f)$, this equation has to be inverted. Based on the WLC model, we tested the influence of the persistence length $P$ in a typical range of 1.0 to $\unit{1.5}{\nano\metre}$ \cite{Dessinges2002}.

\begin{table}[ht]
 \centering \small
\begin{tabular}{|c|c||c|c|c|c||c|c|c|c|}
\hline \multicolumn{2}{|c||}{} & \multicolumn{4}{c||}{molecule 1} & \multicolumn{4}{c|}{molecule 2} \\ \hline \hline
 & parameter & \multicolumn{8}{c|}{change in no.\ of opened bps [bps]} \\ 
 \raisebox{1.5ex}[-1.5ex]{model} & $b$ \& $P$ [nm], $Y$ [pN] & $\Delta n_1$ & $\Delta n_2$ & $\Delta n_3$ & $\Delta n_{\rm{tot}}$ & $\Delta n_1$ & $\Delta n_2$ & $\Delta n_3$ & $\Delta n_{\rm{tot}}$ \\ \hline \hline
    & $b=1.42$, $Y=812$ \cite{Smith1996} & 18 (2)  & 14 (2)  & 15 (2)   & 46 (3) & 13 (1)  & 15 (1)  & 15 (1)   & 42 (2)  \\
\raisebox{1.5ex}[-1.5ex]{FJC}
    & $b=1.15$, $Y= \infty$ \cite{Huguet2010} & 19 (3)   & 15 (2) & 16 (2)  & 50 (3)  & 14 (1) & 16 (1)   & 16 (1)   & 46 (2)  \\ \hline
 & $P=1.0$ & 20 (3)   & 16 (3) & 17 (3) & 54 (3) & 14 (1) & 17 (1) & 17 (1) & 49 (2) \\ 
WLC & $P=1.3$ \cite{Rivetti1998} & 20 (3) & 15 (2) & 16 (2) & 51 (3) & 14 (1) & 16 (1) & 16 (1) & 46 (2) \\ 
    & $P=1.5$ & 19 (3)   & 15 (2)  & 16 (2)  & 50 (3) & 13 (1)  & 16 (1)  & 16 (1)   & 45 (2)   \\ \hline \hline
 \multicolumn{2}{|c||}{expected values} & 21   & 16   & 16   & 53   & 21   & 16   & 16   & 53   \\ \hline 
\end{tabular}
\caption{Overview over the change in the number of opened bps for different models and parameters for the two representative molecules. The numbers in brackets are the standard deviations.}\label{tab: results_n}
\end{table}

From the data shown in table \ref{tab: results_n} it is apparent that, depending on the model and parameters, the results for the estimated change in the number of opened bps vary in an acceptable range when the errors are taken into account. In addition, one can see that it is not evident which model and parameters should be considered as the best ones. Good results are found for the FJC model using recent values of \cite{Huguet2010} and for the WLC model with $P=\unit{1.3}{\nano\metre}$ \cite{Rivetti1998}.
We chose to work with the FJC model with the parameters of \cite{Huguet2010} in sec.~\ref{sec: theory}.

The second molecule indeed reveals a considerably smaller $\Delta n_1$, depending on the model around 13 or $\unit{14}{bps}$, so that 7 or $\unit{8}{bps}$ are not closed after the folding process is completed. Performing single-molecule experiments without knowing the exact influence of permanently frayed bps can lead to misinterpreted results. Checking the appropriate parameters for the polymer models with the help of the change in the number of opened bps of the hairpin branches, one can estimate the number of irreversibly frayed bps at the basis of the stem\footnote{We like to note that the checking of the change in the number of opened bps can be, in principle, also applied to non-permanent, reversible molecular fraying.}.

\section{Theory for the unfolding kinetics}\label{sec: theory}

\begin{figure}[htb]
\begin{center}
\subfigure[]
{\label{fig: transition_rates}\includegraphics[scale=0.53]{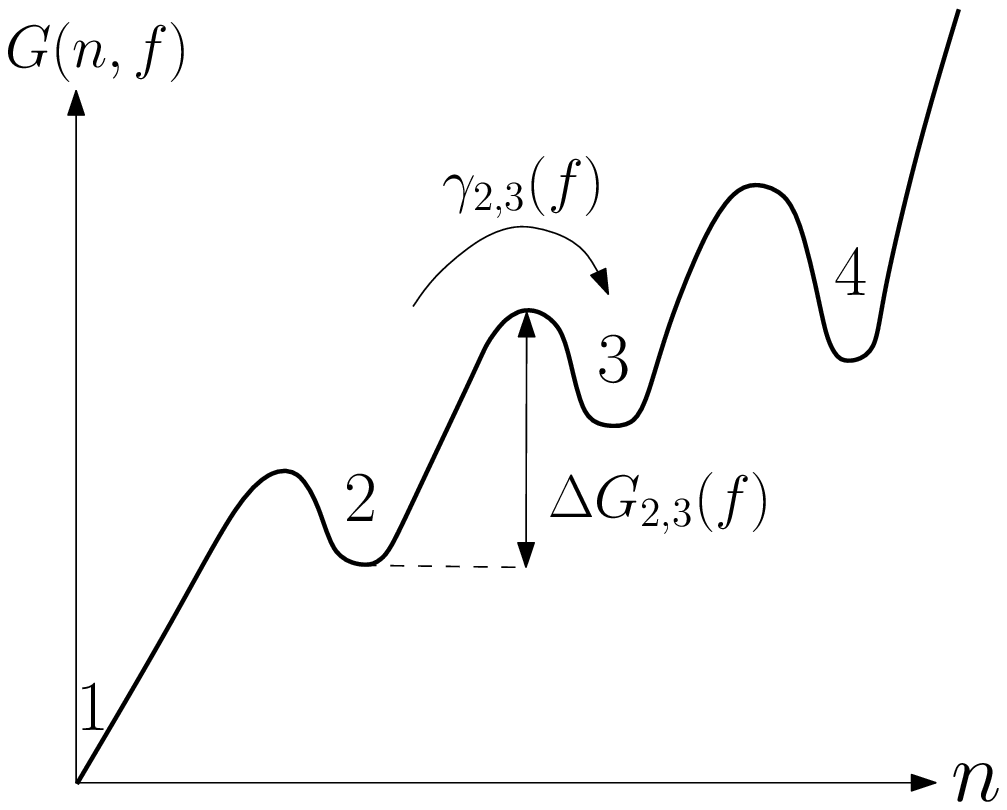}} 
\subfigure[]
{\label{fig: FEL_2forces}\includegraphics[scale=0.3]{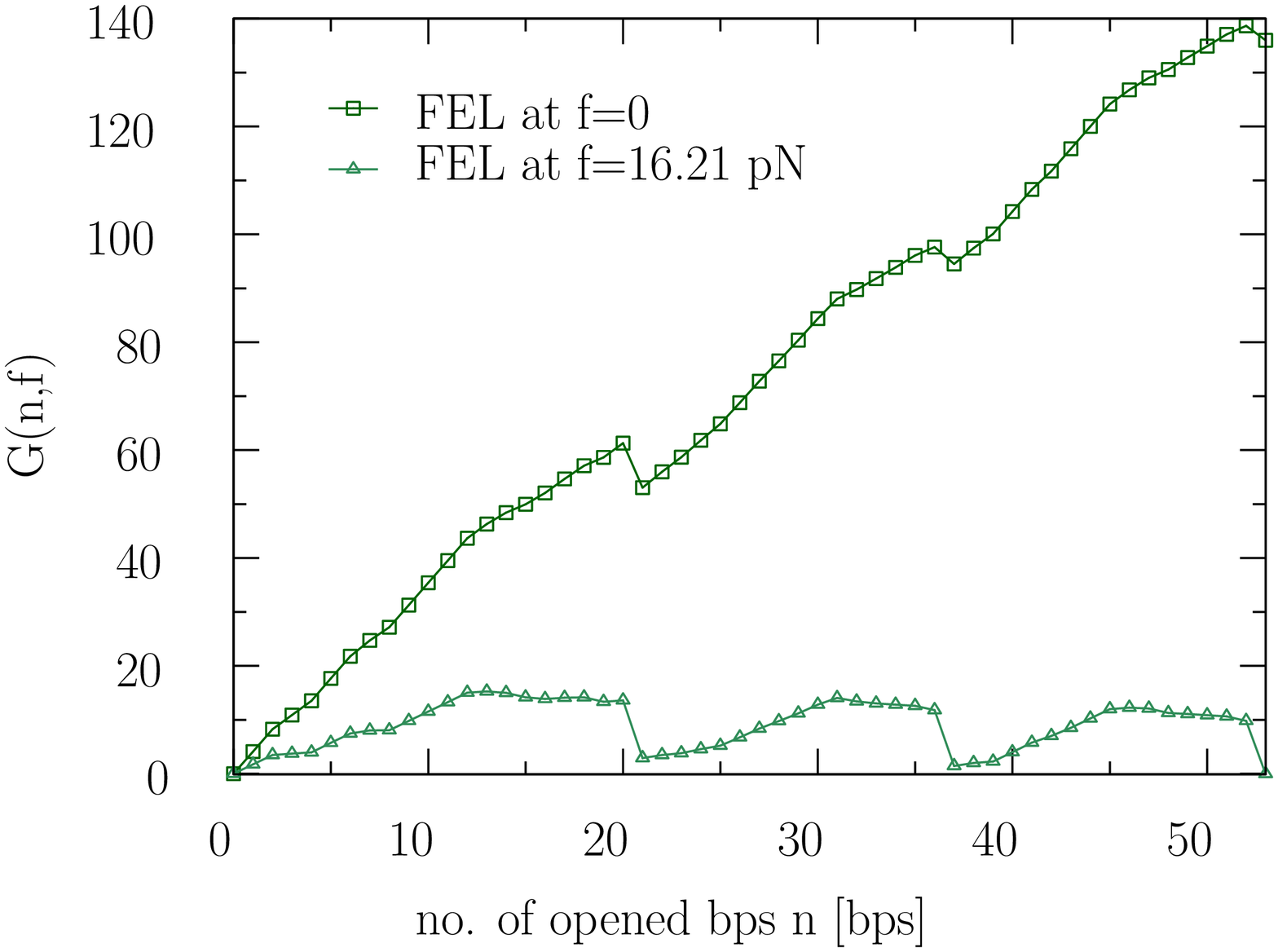}} 
\end{center}
\caption{(a) Sketch of the expected FEL for the triple-branch molecule at zero force as a function of the number $n$ of opened bps. (b) FEL calculated from eqs.~(\ref{eq: g}) to (\ref{eq: g_part}) at two different forces in units of the thermal energy $k_\mathrm{B}T$.}
\label{fig: FEL_graphs}
\end{figure}

The kinetics of the unfolding process can be described on a coarse-grained level based on a Gibbs free energy $G(n,f)$ as a function of the number $n$ of sequentially opened bps for an applied force $f$. 
For small forces, including $f=0$, the FEL is expected to have a shape as displayed in fig.~\ref{fig: transition_rates}. 
In general, $G(n,0)$ increases monotonously with $n$. However, local minima occur at the metastable states 2, 3 and 4 because there is an increase of entropy associated with the release of additional degrees of freedom when the stem-hairpin-junction and end-loops of the hairpin branches are opened.
With rising force the FEL is expected to get tilted, so that the energies of the metastable states are lowered.  
With a knowledge of $G(n,f)$ we can apply standard transition rate theory and write for the transition rate from state $i$ to $i+1$ 
\begin{equation} 
\Gamma_{i,i+1}(f) = \gammaup\smni \; \gammaup_{i,i+1}(f)\,,
\label{eq: gamma}
\end{equation}
where $\gammaup\smni$ is an attempt rate and $\gammaup_{i,i+1}(f)$ is the Boltzmann factor corresponding to the activation barrier $\Delta G_{i,i+1}(f)$ that has to be surmounted,
\begin{equation}
 \gammaup_{i,i+1} (f) = \exp\left(-\frac{\Delta G_{i,i+1}(f)}{k_\mathrm{B}T}\right)\,.
\label{eq: gammai}
\end{equation}

When considering only sequential configurations in the evaluation of the FEL,  different structures compatible with a given $n$ can occur once the stem is completely unfolded. These refer to different possibilities of breaking the bps in the hairpin branches 1 and 2. In order to point out this "degeneration", a new parameter $\alpha$ is introduced, leading to the FEL $G(n,\alpha,f)$. At a given $n$, we must calculate the (restricted) partition sum over the configurations $\alpha$ to get $G(n,f)$.
In order to find $G(n,\alpha,f)$, we consider the following decomposition, 
\begin{equation} 
G(n,\alpha,f)=\Gform+\Gstr -f \dxss\,,
\label{eq: g}
\end{equation}
where $\Gform$ is the free energy of formation of the configuration $(n,\alpha)$, $\Gstr$ is the strain energy of the unfolded ssDNA and $f \dxss$ is a Legendre term (for the definition of $\dxss$ see eq.~(\ref{eq: Gstr}) below).

The free energy of formation $\Gform$ is written as
\begin{equation}
\Gform = \sum_{\rm{all\; bps}} g_{\mu,\mu+1} + G_{\rm junc}(n) + G^{(1)}_{\rm loop}(n) + G^{(2)}_{\rm loop}(n)\,.
\label{eq: gform}
\end{equation}
The first term refers to the nearest neighbour model developed in \cite{Zimm1964, Tinoco1962}, which specifies the interaction $g_{\mu,\mu+1}$ between a base pair $\mu$ and the directly adjacent one $\mu+1$.
It was shown to provide reasonable agreement with experiments \cite{Manosas2006, Mossa2009a, Woodside2006a}.
For example, applying this model onto a sequence $5'$-TCCAG\ldots-$3'$ and its complementary part $3'$-AGGTC\ldots-$5'$, the stack energy reads  $G_{\rm stack}=g_{\rm TC/AG}+g_{\rm CC/GG}+g_{\rm CA/GT}+g_{\rm AG/TC}+\ldots$ 
The most recent values of $g_{\mu,\mu+1}$ lie in the range of -2.37 to \unit{-0.84}{kcal/mol} at 25\textcelsius \, \cite{Huguet2010}.
The terms $G_{\rm junc}(n)$, $G^{(1)}_{\rm loop}(n)$ and $G^{(2)}_{\rm loop}(n)$ in eq.~(\ref{eq: gform}) describe the free energy reduction due to the release of the stem-hairpin-junction and end-loops and are estimated from \cite{SantaLucia1998,Zuker2003} as $G^{(1)}_{\rm loop} = G^{(2)}_{\rm loop} = \unit{1.58}{kcal/mol}$ and $G_{\rm junc} = \unit{4.90}{kcal/mol}$.

The strain energy $\Gstr$ of the unfolded single-stranded part \cite{Mossa2009a} with contour length $l=l(n,\alpha)$\footnote{In previous publications of some of the authors this contour length was denoted by $l_{n,\alpha}$ to emphasize the dependence on $n$ (and, in addition, $\alpha$ here). This dependence is caused by the change of the contour length in the transitions. For easier reading we suppress to give it explicitely in the following. Further details about the contour length were already discussed in sec.~\ref{sec: patterns}.} can be calculated from the work needed to stretch the unpaired bases.
We like to remind the reader that we denote the elongation of the mean end-to-end distance of the ssDNA in force direction by $u_l(f)$. In what follows we chose to work with the FJC model, see eq.~(\ref{eq: FJC}), with parameters of \cite{Huguet2010} instead of using the WLC model, despite the fact that both approaches give similar good results (cp.~table \ref{tab: results_n}).
Since $u_l(f)$ is monotonously increasing with $f$, it has an inverse $f_l(u) = u_l^{-1}(f)$, which is the force that is exerted by a ssDNA chain with contour length $l$, if its mean end-to-end distance is elongated by $u$.
Accordingly, setting $\dxss = u_l(f)$, we can write 
\begin{equation}
\Gstr = \int \limits_{0}^{\dxss}  du' \, f_l(u') = f\dxss - \int \limits_0^{f} df' \, u_l(f')\,.
\label{eq: Gstr}
\end{equation}

Finally, we computed $G(n,f)$ by
\begin{equation}
G(n,f) = -k_\mathrm{B} T \ln \sum_\alpha \exp\left(-\frac{G(n,\alpha,f)}{k_\mathrm{B}T}\right)\,.
\label{eq: g_part}
\end{equation}

In fig.~\ref{fig: FEL_2forces} the FEL is depicted for $f=0$ and $f=\unit{16.21}{\pico\newton}$. It exhibits the behaviour anticipated in fig.~\ref{fig: transition_rates}: for zero force it has minima at the stable/metastable states and it becomes tilted with rising force. When approaching the force regime where the rips occur in fig.~\ref{fig: u_molA}, the levels of the minima become comparable.
We want to point out that the DNA sequences shown in fig.~\ref{fig: 3helixmol}  have been designed deliberately to yield the multiple-state structure seen in fig.~\ref{fig: FEL_2forces}. This gives us some confidence in the model underlying the construction of the $G(n,\alpha,f)$ in eq.~(\ref{eq: g}).

Based on the FEL we can easily calculate the transition probability $W(f_i|f_{i-1})$ for the first rupture force $f_i$ in the $i$th transition if the first rupture force was $f_{i-1}$ in the $(i-1)$th transition. The result is 
\begin{equation}
W(f_i|f_{i-1})
= \frac{\gammaup^{\scriptscriptstyle{0}}_{\scriptscriptstyle{i}}}{r} \; \gammaup_{{i,i+1}}(f) \; \exp\Big[{- \frac{\gammaup^{\scriptscriptstyle{0}}_{\scriptscriptstyle{i}}}{r} 
\int_{f_{i-1}^*}^{f}df'\;\gammaup_
{{i,i+1}}(f')}\Big]\,,
\end{equation}
where $f_{i}^*=f_i - \overline{\Delta f_i}$ (and $f_0=f^*_0=0$); $r$ was the loading rate, see sec.~\ref{sec: exp}, and $\gammaup\smni$ and $\gammaup_{i,i+1}(f)$ were defined in eq.~(\ref{eq: gamma}).
The activation energy $\Delta G_{i,i+1}(f)$ appearing in eq.~(\ref{eq: gammai}) was calculated, as indicated in fig.~\ref{fig: transition_rates}, from the $G(n,f)$ by determining the energy $G^{\rm{min}}_i(f)$ of the local minimum belonging to state $i$ and the saddle point energy $G^{\rm{saddle}}_{i,i+1}(f)$ of the $i$th transition between the $i$th and $(i+1)$th state, $\Delta G_{i,i+1}(f) = G^{\rm{saddle}}_{i,i+1}(f) - G^{\rm{min}}_i(f)$. The attempt rate $\gammaup\smni$ was used as the only fitting parameter.

For the joint probability density of the three first rupture forces we then obtain 
\begin{equation}
\Psi_3(f_1,f_2,f_3) = W(f_1|0) \;W(f_2|f_1) \;W(f_3|f_2)\,,
\end{equation}
which allows us to calculate the distributions shown in fig.~\ref{fig: histo_frf_fj_molA}.

\begin{figure}[htb]
\begin{center}
\includegraphics[scale=0.4,angle=-90]{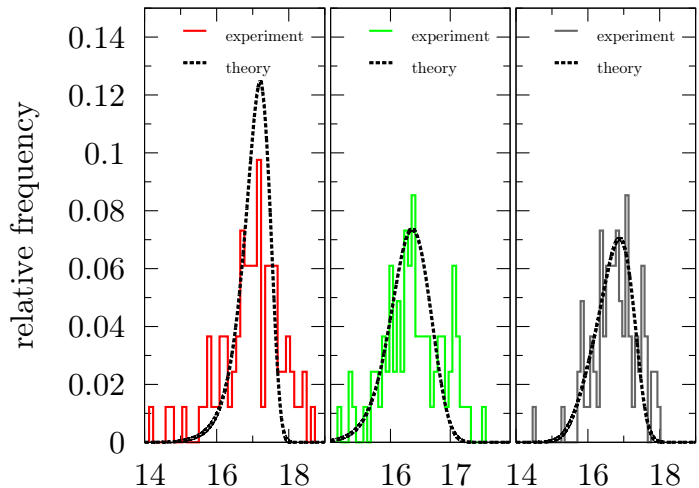}
\end{center}
\caption{Comparison of the first rupture force distributions from fig.~\ref{fig: histo_frf_fj_molA} with the theory for the first ($f_1$), the second ($f_2$) and the third ($f_3$) transition, see left, middle and right panel, respectively. The best fit was obtained for attempt rates  $\gammaup^{\scriptscriptstyle{0}}_{\scriptscriptstyle{1}} = \unit{9\cdot 10^6}{Hz}$, $\gammaup^{\scriptscriptstyle{0}}_{\scriptscriptstyle{2}} = \unit{1.1\cdot 10^6}{Hz}$ and  $\gammaup^{\scriptscriptstyle{0}}_{\scriptscriptstyle{3}} = \unit{3.5\cdot 10^5}{Hz}$.}
\label{fig: theory_exp}
\end{figure}

Figure \ref{fig: theory_exp} displays the histograms for the three rips from fig.~\ref{fig: histo_frf_fj_molA} in comparison with the distributions calculated from our theory. In view of the available statistics (82 cycles, see sec.~\ref{sec: exp}), the agreement is quite satisfactory.

\section{Conclusions}

An important class of biophysical studies is the investigation of junctions in molecules since they present manifold ways to interact with other substances, for instance cations. Three-way junctions are especially interesting because metal ions such as magnesium \cite{Liphardt2001, Manosas2005} can bind to them and alter the tertiary structure.
Here a first step of such a study is presented where we investigate a molecule with a three-way junction alone, without in vivo relevant substances.

One of our aims of this work was to study whether the construction and kinetics of more complex DNA molecules with richer folding-unfolding behaviour can be described by proper extensions of theories developed successfully for two-state systems so far. Our results show that this is indeed possible, at least for the unfolding trajectories. 
A triple-branch molecule has been specifically designed to produce a four-state system based on a model for the free energy landscape. This design was successful and we were able to prove the existence of these states by the emergence of associated force rips in pulling experiments. 
The first rupture forces have been systematically recorded in these pulling experiments and their distributions have been calculated. 
A transition rate theory based on the free energy landscape was successful in describing these distributions. 

Two patterns have been found in the measured unfolding trajectories, one indicating the anticipated unfolding behaviour and the other one pointing to the occurrence of irreversible molecular fraying. This characterisation was possible by connecting the extracted force jump values to the change in the number of opened bps at each transition.
For this estimation we 
tested the validity of two polymer models for the elastic response of ssDNA (FJC and WLC) and different sets of parameters in order to find the best agreement with the expected values. This analysis is useful to compare the elastic properties measured in DNA unzipping experiments with those obtained by stretching ssDNA polymers \cite{Huguet2010}.

One class of molecules required a smaller force than anticipated to unfold the stem since some bps at its basis are partly or completely melted due to photodamaging.
Permanent molecular fraying is an usually undesired, but frequent effect in single-molecule studies and deserves special attention in order to reduce its distorting influence on experimental results.
It is important to find means to avoid irreversible fraying since not fully closed molecules change the measured unfolding-folding trajectories so that an average over all molecules, including permanently frayed ones, can lead to deviations of the real values and to misinterpreted results. To improve the statistics of the results, it is therefore necessary to identify irreversibly frayed molecules and remove them from the analysis. Within our analysis, we found a useful method to identify permanent molecular fraying. With the appropriate parameters for the polymer models one can estimate the number of irreversibly frayed bps at the basis of the stem.
Further experiments performed with silica beads instead of polystyrene microspheres could clarify under which conditions permanent molecular fraying can be decreased.

All this knowledge paves the way to further interesting studies such as the refolding problem \cite{Manosas2008} or two topics which require further experimental research. 
Firstly, the binding of metal ions to the three-way junctions could be examined in order to find out about structural changes due to the formation of tertiary contacts and therefore altered kinetics of the unfolding process \cite{Manosas2005}. 
Secondly, the translocation motion of helicases that unwind dsDNA could be addressed as another interesting subject,  
including the investigation of how they move along bifurcation points. Eventually, the kinetic approach for the prediction of the probability distributions of the first rupture forces could be a well suited starting point for future in-depth modelling in this domain on a more microscopic basis.

\section*{Acknowledgements}

S.~E.\ thanks the Deutscher Akademischer Austauschdienst (DAAD) for providing financial support (FREE MOVER and PROMOS)  for stays at the Small Biosystems Lab in Barcelona where the experiments have been performed.
A.~A.\ is supported by grant AP2007-00995. F.~R.\ is supported by the
grants FIS2007-3454, Icrea Academia 2008 and HFSP (RGP55-2008).

\bibliographystyle{jbact}   
\bibliography{references}  

\begin{thebibliography}{}

\bibitem{Ammenti2009}
{\bf Ammenti, A., F.~Cecconi, U.~M.~B. Marconi, and A.~Vulpiani.}  2009.
\newblock J Phys Chem B  {\bf 113} (30):10348--10356.

\bibitem{Block2003}
{\bf Block, S.~M., C.~L. Asbury, J.~W. Shaevitz, and M.~J. Lang.}  2003.
\newblock Proc Natl Acad Sci U S A  {\bf 100} (5):2351--2356.

\bibitem{Bustamante1994a}
{\bf Bustamante, C., J.~F. Marko, E.~D. Siggia, and S.~Smith.}  1994.
\newblock Science  {\bf 265} (5178):1599--1600.

\bibitem{Carter2006}
{\bf Carter, N.~J., and R.~A. Cross.}  2006.
\newblock Curr Opin Cell Biol  {\bf 18} (1):61--67.

\bibitem{Zimm1964}
{\bf Crothers, D.~M., and B.~H. Zimm.}  1964.
\newblock J Mol Biol  {\bf 9}:1--9.

\bibitem{Dessinges2002}
{\bf Dessinges, M.-N., B.~Maier, Y.~Zhang, M.~Peliti, D.~Bensimon, and
  V.~Croquette.}  2002.
\newblock Phys Rev Lett  {\bf 89} (24):248102.

\bibitem{Tinoco1962}
{\bf Devoe, H., and I.~J. Tinoco.}  1962.
\newblock J Mol Biol  {\bf 4}:500--517.

\bibitem{Dumont2006}
{\bf Dumont, S., W.~Cheng, V.~Serebrov, R.~K. Beran, I.~Tinoco, A.~M. Pyle, and
  C.~Bustamante.}  2006.
\newblock Nature  {\bf 439} (7072):105--108.

\bibitem{Forns2011}
{\bf Forns, N., S.~{de Lorenzo}, M.~Manosas, K.~Hayashi, J.~M. Huguet, and
  F.~Ritort.}  2011.
\newblock Biophys J  {\bf 100} (7):1765--1774.

\bibitem{Hormeno2006}
{\bf Hormeno, S., and J.~R. Arias-Gonzalez.}  2006.
\newblock Biol Cell  {\bf 98} (12):679--695.

\bibitem{Huguet2010}
{\bf Huguet, J.~M., C.~V. Bizarro, N.~Forns, S.~B. Smith, C.~Bustamante, and
  F.~Ritort.}  2010.
\newblock Proc Natl Acad Sci U S A  {\bf 107} (35):15431--15436.

\bibitem{Landry2009}
{\bf Landry, M.~P., P.~M. McCall, Z.~Qi, and Y.~R. Chemla.}  2009.
\newblock Biophys J  {\bf 97} (8):2128--2136.

\bibitem{Leger1998}
{\bf Leger, J.~F., J.~Robert, L.~Bourdieu, D.~Chatenay, and J.~F. Marko.}
  1998.
\newblock Proc Natl Acad Sci U S A  {\bf 95} (21):12295--12299.

\bibitem{Liphardt2001}
{\bf Liphardt, J., B.~Onoa, S.~B. Smith, I.~Tinoco, and C.~Bustamante.}  2001.
\newblock Science  {\bf 292} (5517):733--737.

\bibitem{Maier2000}
{\bf Maier, B., D.~Bensimon, and V.~Croquette.}  2000.
\newblock Proc Natl Acad Sci U S A  {\bf 97} (22):12002--12007.

\bibitem{Manosas2006}
{\bf Manosas, M., D.~Collin, and F.~Ritort.}  2006.
\newblock Phys Rev Lett  {\bf 96} (21):218301.

\bibitem{Manosas2008}
{\bf Manosas, M., I.~Junier, and F.~Ritort.}  2008.
\newblock Phys Rev E Stat Nonlin Soft Matter Phys  {\bf 78} (6 Pt 1):061925.

\bibitem{Manosas2005}
{\bf Manosas, M., and F.~Ritort.}  2005.
\newblock Biophys J  {\bf 88} (5):3224--3242.

\bibitem{Melchionna2007}
{\bf Melchionna, S., M.~Fyta, E.~Kaxiras, and S.~Succi.}  2007.
\newblock Int. J. Mod. Phys. C  {\bf 18}:685.

\bibitem{Meller2000}
{\bf Meller, A., L.~Nivon, E.~Brandin, J.~Golovchenko, and D.~Branton.}  2000.
\newblock Proc Natl Acad Sci U S A  {\bf 97} (3):1079--1084.

\bibitem{Mossa2009a}
{\bf Mossa, A., M.~Manosas, N.~Forns, J.~M. Huguet, and F.~Ritort.}  2009.
\newblock J. Stat. Mech. {\bf P02060}.

\bibitem{Ritort2006}
{\bf Ritort, F.}  2006.
\newblock J. Phys. Condens. Matter  {\bf 18}:R531--R583.

\bibitem{Ritort2006a}
{\bf Ritort, F., S.~Mihardja, S.~B. Smith, and C.~Bustamante.}  2006.
\newblock Phys Rev Lett  {\bf 96} (11):118301.

\bibitem{Rivetti1998}
{\bf Rivetti, C., C.~Walker, and C.~Bustamante.}  1998.
\newblock J Mol Biol  {\bf 280} (1):41--59.

\bibitem{SantaLucia1998}
{\bf SantaLucia, J.~J.}  1998.
\newblock Proc. Natl. Acad. Sci. USA  {\bf 95}:1460--1465.

\bibitem{Smith2001}
{\bf Smith, D.~E., S.~J. Tans, S.~B. Smith, S.~Grimes, D.~L. Anderson, and
  C.~Bustamante.}  2001.
\newblock Nature  {\bf 413} (6857):748--752.

\bibitem{Smith1996}
{\bf Smith, S.~B., Y.~Cui, and C.~Bustamante.}  1996.
\newblock Science  {\bf 271} (5250):795--799.

\bibitem{Smith2003}
{\bf Smith, S.~B., Y.~Cui, and C.~Bustamante.}  2003.
\newblock Methods in Enzymology  {\bf 361}:134--162.

\bibitem{Mameren2009}
{\bf {van Mameren}, J., P.~Gross, G.~Farge, P.~Hooijman, M.~Modesti,
  M.~Falkenberg, G.~J.~L. Wuite, and E.~J.~G. Peterman.}  2009.
\newblock Proc Natl Acad Sci U S A  {\bf 106} (43):18231--18236.

\bibitem{Wagner2006}
{\bf Wagner, B., R.~Tharmann, I.~Haase, M.~Fischer, and A.~R. Bausch.}  2006.
\newblock Proc Natl Acad Sci U S A  {\bf 103} (38):13974--13978.

\bibitem{Wang1997}
{\bf Wang, M.~D., H.~Yin, R.~Landick, J.~Gelles, and S.~M. Block.}  1997.
\newblock Biophys J  {\bf 72} (3):1335--1346.

\bibitem{Wen2008}
{\bf Wen, J.-D., L.~Lancaster, C.~Hodges, A.-C. Zeri, S.~H. Yoshimura, H.~F.
  Noller, C.~Bustamante, and I.~Tinoco.}  2008.
\newblock Nature  {\bf 452} (7187):598--603.

\bibitem{Woodside2006a}
{\bf Woodside, M.~T., W.~M. Behnke-Parks, K.~Larizadeh, K.~Travers,
  D.~Herschlag, and S.~M. Block.}  2006.
\newblock Proc Natl Acad Sci U S A  {\bf 103} (16):6190--6195.

\bibitem{Wuite2000}
{\bf Wuite, G.~J., S.~B. Smith, M.~Young, D.~Keller, and C.~Bustamante.}  2000.
\newblock Nature  {\bf 404} (6773):103--106.

\bibitem{Yasuda2001}
{\bf Yasuda, R., H.~Noji, M.~Yoshida, K.~Kinosita, and H.~Itoh.}  2001.
\newblock Nature  {\bf 410} (6831):898--904.

\bibitem{Yin1995}
{\bf Yin, H., M.~D. Wang, K.~Svoboda, R.~Landick, S.~M. Block, and J.~Gelles.}
  1995.
\newblock Science  {\bf 270} (5242):1653--1657.

\bibitem{Zuker2003}
{\bf Zuker, M.}  2003.
\newblock Nucleic Acids Res. {\bf 31} (13):3406--3415.
\newblock See also \url{http://mfold.rna.albany.edu/}.

\end{thebibliography}

\end{document}